# Is ChatGPT as reliable as individual reviewers assessing the quality of published journal articles from PDFs or titles and abstracts?

Mike Thelwall
School of Information, Journalism and Communication, University of Sheffield, UK. https://orcid.org/0000-0001-6065-205X

Parveen Ali
School of Allied Health Professions, Nursing and Midwifery, University of Sheffield, UK. https://orcid.org/0000-0002-7839-8130

Whilst Large Language Models (LLMs) have a weak to moderate ability to score published journal articles for research quality, they have not been compared with individual expert reviewers. It is also unknown whether quality scores from ChatGPT based on PDFs can improve on those from titles and abstracts through a deeper evaluation. To address both issues, this article uses expert scores (98 internal departmental ratings for UK Unit of Assessment [UoA] 3 Allied Health Professions, 44 for UoA13 Architecture, and 58 library and information science articles from UoA34), comparing them against ChatGPT-5.4 scores from both title/abstract and PDF inputs. For UoA3, individual reviewer scores were also compared against each other and ChatGPT-5.4. The rank correlations with expert scores are almost all statistically significantly positive, but differences between the correlations are mostly not, despite weakly suggesting that ChatGPT-5.4 can be more reliable than individual reviewers for UoA3. Moreover, whilst ChatGPT-5.4 provides more detailed evaluations of PDFs than of titles/abstracts, its score predictions do not seem to improve. Thus, whilst the results broadly confirm the value of ChatGPT scores for ranking academic documents, its apparently deeper evaluative comments on PDFs are misleading in the sense of not translating to improved score predictions.

## 1 Introduction

The rise of generative AI (GenAI) has reduced the barrier to academic publishing, particularly for researchers whose first language is not English. This has increased the volume of journal submissions (He & Bu, 2026), and a similar increase seems to have occurred for other academic documents, such as grant applications (Kolarz, 2026) and conference submissions (Choi et al., 2026), increasing the overall burden of peer review. This and the general potential of GenAI have motivated attempts to support or replace academic reviewing tasks with AI (e.g., Carbonell Cortés et al., 2024). In support of this there have been extensive efforts to compare peer review suggestions or recommendations with those of LLMs (e.g., Li et al., 2025; Thakkar et al., 2025), for example, finding that LLMs lack deeper insights into methods errors and field aspirations, whereas humans are more likely to be less comprehensive in commenting. In parallel, a separate strand of research has investigated the ability of LLMs to make an overall numerical quality assessment of academic documents, including conference and journal submissions, published journal articles, and funding applications (Sikimić & Radovanović, 2022; Thelwall, 2026a; Thelwall & Mohammadi, 2026; Thelwall & Yaghi, 2025; Zhao et al., 2025). These have tended to find a weak to moderate degree of agreement between humans and LLMs, as reviewed in more detail below, or between LLMs and citations or downloads (Zhu et al., 2026a). From these, it seems clear that current LLMs are not good enough to entirely replace humans when overall quality assessments are needed for academic documents, but it is not known how their accuracy compares to that of individual experts, and therefore whether it would be reasonable to replace individual human scores with LLM scores.

Previous research using LLMs to allocate overall ratings to academic documents has found that LLM scores are typically inaccurate because they allocate a much narrower range, such as usually giving a high but not maximum score and almost never giving low scores (Thelwall & Yaghi, 2025). In particular, they are lenient on weak papers (Sharma et al., 2026). Nevertheless, whilst LLM scores tend to be inaccurate, their rank order correlates weakly to moderately with the rank order of expert scores for conference papers (Thelwall & Yaghi, 2025; Thelwall, 2026a), journal articles (Thelwall, 2026b) and grant proposals (Sandstrom & Thelwall, 2026). Thus, the appropriate way to use LLM scores is to convert them into ranks and then either use the ranks for decision making or map the ranks onto numerical scores, such as through a lookup table designed to give a human-like overall score distribution.

In terms of individual reviewer and LLM consistency, LLMs tend to agree with each other on raw scores (Baumann et al., 2026), whereas human reviewers have a low degree of agreement with each other (Bornmann et al., 2010; Feliciani et al., 2022; Jirschitzka et al., 2017; Lee et al., 2013) and LLMs have weak agreement with humans on raw scores (Li et al., 2026). Better levels of raw agreement for ChatGPT-5 and Claude-Sonnet-4.5 have been found for an experimental setup with multiple reviewers, but without comparing reviewers against each other (Liu et al., 2026). Previous human-LLM agreement studies have either compared LLMs with a single expert reviewer (e.g., Thelwall, 2025b) or with the average scores of a set of reviewers (e.g., Loc et al., 2026; Li et al., 2026), neither of which give insights into the relative strength of LLM ratings compared to individual reviewers. The only exception is that one small sample study of grant proposals found that the maximum level of agreement (Spearman correlation) between LLMs and reviewers was half as high as the level of agreement between reviewers (Sandstrom & Thelwall, 2026).

A separate issue for the same task is that, with few exceptions, previous investigations using LLMs to score journal articles for quality have used titles and abstracts as inputs (e.g., Thelwall & Yaghi, 2025) or plain text files with tables and figures removed (e.g., Thelwall, 2025b). This has been due to the lack of a PDF processing capability for the major open weights LLMs and in the main LLM APIs. The one exception, Gemini, has not given improved results with PDFs in a small sample analysis (Thelwall, 2025a). The ChatGPT API introduced a PDF processing ability in March 2025, and since ChatGPT has produced the best research quality score predictions so far, it is important to assess whether it can produce better results with PDFs. In addition, it would be useful to know whether its reports contain more specific evaluative components or mention visual elements when the input is the article PDF rather than just the title and abstract.

As motivated above, this article has two separate goals. First, it assesses the relative strength of individual human expert reviewers against that of LLMs for scoring journal articles. Second, it investigates whether ChatGPT can give better results with PDF input than with titles and abstracts. Although these issues are relevant to many types of academic document, for practical reasons the scope here is quality assessments of published journal articles. Since there are now thousands of LLMs and their power of the newest releases seems to increase monthly, any study is dealing with a moving target. Nevertheless, investigating this issue for current highest performing LLM, ChatGPT-5, at least give a benchmark for what is possible to inform applications and so that future new LLMs can be compared. The research questions are as follows.

- RQ1: Are ChatGPT-5 score ranks as accurate as individual human score ranks for published journal articles (for UoA3)?
- RQ2: Can ChatGPT-5 predict article quality from PDFs better than from titles and abstracts (for all three UoAs)?
- RQ3: Do ChatGPT-5 reports on PDFs differ from its reports on titles and abstracts (for all three UoAs)?

# 2 Methods

The research design was to obtain sets of academic journal articles from different fields with multiple expert reviewer scores, then obtain ChatGPT-5 scores for the articles and compare the ChatGPT-5 and human score ranks against each other, with both title/abstract and PDF inputs.

## *2.1 Data: Expert scores*

Scoring the overall quality of academic research normally requires an experienced researcher from the same field, if not the same speciality. In the UK, research outputs are scored nationally for research quality every seven years, each by at least two experts (mostly full professors) from a panel of 1100 as part of the Research Excellence Framework (REF). This process allocates both funding and reputation. These scores (e.g., those from REF2021) would be ideal for this study but are not published. Every year, UK universities conduct their own internal mini-REFs as part of their preparation for the national REF; these scores are confidential and never published. They are used to rank the publications produced by each department so that the top n can be submitted to the next REF (REF2029), where n = 2.5 times the number of full-time equivalent researchers in the relevant department. The mini-REF scores are close to ideal for the current study since they are produced by independent field experts for the important purpose of REF2029 article selection. Since they are confidential, these scores have the advantage of not being part of any LLM training data, ruling out the possibility of cheating in experiments.

Accessing private departmental mini-REF scores is a sensitive issue because the scores may affect academic careers and REF2029 outcomes if made public. It is therefore important to approach data access ethically and sensitively. The University of Sheffield was chosen as the source of the mini-REF data to be requested due to author affiliation with the university, given that the data request would put a substantial burden on each department as well as trust about the confidentiality of the data. The original plan was to approach four UoAs, one in each of the four main REF main panels: A health and life sciences, B physical sciences and engineering, C social sciences, and D arts and humanities. First, ethical approval was obtained. Next, the University REF lead was approached and permission sought from the University REF team and the unions. Next UoA leads in the author’s department (main panel D) and the largest UoAs in each other main panel were contacted for permission. The largest UoAs were selected to give the greatest statistical power in testing. In two cases, this was denied so the lead for the next largest UoA was approached as a replacement. For Main Panel B, agreement in principle was reached for two UoAs (one who volunteered to collaborate) but did not progress despite several meetings (apparently due to understandable internal opposition from other members of the departmental REF teams). We eventually obtained authorisation from UoA 3 (Allied Health Professions, Dentistry, Nursing and Pharmacy), UoA 13 (Architecture, Built Environment and Planning), and UoA 34 (Communication, Cultural and Media Studies, Library and Information Management) to survey researchers for approval to see their mini-REF scores for one year. We then obtained the allowed scores (i.e., for researchers that had given permission) from the relevant UoA administrator or REF lead. For UoA 3, we were given individual reviewer scores but for UoA 13 and 34 we were only given the REF panel team agreed score for each output (usually the average of the individual reviewer scores, unless overridden by the team), which was not useful for RQ1. The raw data for the current project therefore includes final scores for three UoAs and individual scores for one. The permission seeking process took about 18 months altogether. The UoA 34 dataset has been analysed previously with ChatGPT-4o and Gemini 2.0 (Langfeldt et al., 2026).

The raw data consists of the identities of journal articles and between two and five expert reviewer scores (UoA 3) and departmental agreed scores (UoAs 3, 13, 34) for each article (Table 1). All mini-REF reviewers must be at least at senior lecturer level in the University of Sheffield. For UoA 3, there are two constituent departments, and one reviewer was taken from each department for every article so there would usually be one specialist and one non-specialist reviewer. For all three UoAs, additional reviewers were sought when the original two reviewers substantially disagreed or the article score was on the (ranked) borderline of articles potentially submittable to REF2029. These additional reviewers were ignored for the inter-reviewer comparisons since they selected based on reviewer agreement, so would bias the results.

Table 1: The three datasets used.

| UoA | Articles | Individual reviewer scores | Agreed scores | RQs |
|---|---|---|---|---|
| 3: Allied Health Professions, Dentistry, Nursing and Pharmacy | 98 | Y | Y | 1,2,3 |
| 13: Architecture, Built Environment and Planning | 44 | N | Y | 2,3 |
| 34: Communication, Cultural and Media Studies, Library and Information Management | 58 | N | Y | 2,3 |

## 2.2 Data: Journal article versions

The journal articles from all three UoAs were downloaded in PDF form and edited to remove all metadata, creating an anonymous PDF. This included removing author names and affiliations, journal names, citation details, advertisements, page numbers and headers and footers. This just left the title, abstract, main body and references, including any figures, tables and equations. This is therefore adequate for an anonymous full evaluation. Removing all document metadata reduces the chance of the scores being influenced by anything other than the article text.

As an additional input, the PDFs were converted to plain text, removing figures, tables, and references to focus on the main body and to allow processing via text-only submission, as previously used with ChatGPT. Tables were removed to make a uniform approach to aid in interpretation of the results because some tables could not be converted to text in a reasonable way, especially those with columns spanning multiple rows. Figure and table captions were retained. A dataset with just articles and titles was also created for RQs 2 and 3.

## *2.3 ChatGPT 5.4 scores*

Although there are currently thousands of LLMs and ideally the performance of all should be evaluated this is not practical and, because of the small sample sizes, would probably lead to good results by chance from weak models. Thus, the strategy was instead to pre-select the latest version of a LLM that has shown the best performance overall on this type of task. Based on prior research (Thelwall, 2026b), the best performing LLM is ChatGPT 5. The latest version at the time of data collection (June 2026) was 5.5 but version 5.4 was used instead to allow different size models to be tested and compared: Only one version of ChatGPT 5.5 was available. The difference between the two versions may be that ChatGPT 5.5 is ChatGPT 5.4 with some additional training, rather than it being a completely new model.

All three main versions of ChatGPT 5.4 were used because the considerable cost difference between them suggests that it may be useful to know whether the expenditure on larger models could be justified by increased performance. The default API parameters were used since previous research had found them to work at least as well as alternatives (Thelwall, 2025b). The reasoning parameter was set to “none” since prior research had found that reasoning did not improve scores (Thelwall & Mohammadi, 2026).

The models were guided by both system prompts and user prompts, unchanged from prior studies (see: Thelwall, 2026b). The system prompts (see Appendix) are slightly modified versions of the human reviewer instructions for REF2021 (REF2021, 2019) so represent the definitive task description. The Main Panel A instructions were used for UoA 3, the Main Panel C instructions for UoA 13 and the Main Panel D instructions for UoA 34. The sole modification was adding an initial sentence to introduce the task to the LLM. The system prompts in both cases were quite similar, describing the task, the scoring system (1* to 4) and the scoring criteria (rigour, originality and significance).

In contrast, the user prompts differed from the national instructions (which require a single integer score for each paper and no commentary) and instead mimic the University of Sheffield internal score process, which requires separate scores for rigour, originality and significance as well as an overall score and justifications for each the four scores, as well as a 13-point scoring system, 1*, 1.25*, 1.5*,… 4*. The Sheffield processes apparently mimic the internal REF2021 processes, which also used a finer-grained scoring before switching to the integer results, with both UoA 3 (REF2021, 2022a), and UoA 13 (REF2021, 2022b) using a 13-point scale, although no intermediate scale has been officially declared for UoA 34 (REF2021, 2022c).

Following testing on a different dataset (Thelwall, 2026b), the Sheffield instruction ordering was swapped to ask for a score justification before the score. Also, instead of requesting quarter scores, the model was prompted that fractional scores were acceptable. This is a weak type of chain-of-thought or reasoning approach since the model is guided to "think" through each dimension before scoring and the same for the final overall score. This structured user prompt was as follows (Thelwall, 2026b), with the title/abstract or full text appended afterwards into the user prompt or flagged as attached.

> Score this article in the range 1* to 4*, including fractions. Use the following format.
> Rigour report:
> Rigour score:
> Originality report:
> Originality score:
> Significance report:
> Significance score:
> Overall report:
> Overall score:
> ###

As mentioned above, previous research using LLMs for article scoring has shown that the individual scores are inaccurate and the most useful results can be obtained by submitting the query multiple times, averaging the scores, and then ranking the average scores (Thelwall, 2026b). This approach is necessary because of the individual scores are highly inaccurate and clustered around 3, but there is a tendency for LLMs to occasionally give higher scores to better articles and lower scores to worse articles. A slight improvement can be made if the wording of the prompt is changed each time (Thelwall, 2026c) and so five small variants of the above prompt were used, each slightly altering the wording of the first sentence. The sentences below are the four alternatives to the one starting the user prompt above, with all prompts submitted six times to give a total of 30 score predictions per article.

- Score this article between 1* and 4*, including fractions.
- Score this article from 1* to 4*, including fractions.
- Score this article in the range 1* to 4*, including fractions if necessary.
- Score this article from 1* to 4*, including fractions when needed.

The final ChatGPT score for each article and input type (i.e., title/abstract, full text, PDF) was the arithmetic mean of the 30 scores returned by the ChatGPT API from the five prompts submitted six times each. Since three different versions of ChatGPT were tested (i.e., full, mini, nano), nine averages of 30 scores were calculated for each article, one per version and input type.

### *2.4 Analysis*

For RQ1, the relative accuracy of LLMs and human expert scores cannot be calculated because there is no known "true" quality score for any article (including those in the current datasets). In the absence of this, the best approximation to a true score in practice is the average score of multiple independent experts, perhaps after mediation by another panel of experts (as in REF2021). At a more theoretical level, the quality of an article could be argued to be the average score of *all* researchers within the article's specialty. In any case the departmental agreed score (the reviewer average, after departmental REF team mediation) is the best available proxy for article true quality scores. For UoA 3, this was not available and the reviewer average score was used instead (including any additional reviewers after the first two). For RQ1, it is not fair to compare the individual human scores to this departmental average score (because they are component parts of it) so the only reasonable comparison is of the individual human and LLM scores against each other. As mentioned above, only the first two reviewer scores for each article can be used because subsequent scores were only obtained for difficult or marginal cases.

Interpreting the results of correlations between human and LLM scores requires prior hypotheses because of the absence of true scores. In particular, it must be assumed that the human ranks are approximations of the “true” ranks with the errors being due to random noise rather than systematic bias. With this assumption, if an LLM has no ability to rank articles then its correlation with the individual human score ranks would be 0. If it has an exactly human-like ability to rank articles, then its correlations with individual sets of human scores should be identical to the correlations between two different sets of human scores. If its ability to rank articles is weaker than that of humans, then its correlation with individual sets of human scores should be less than the rank correlation between two sets of human scores. Thus, the LLM “fraction” of human level expertise can be approximated by dividing he LLM-human correlation by the human-human correlation. Conversely, if the LLM is more powerful than a single human, then it can correlate more strongly with individual human scores that the human scores correlate with each other. Again, the same fraction (this time greater than 1) would reflect this. Since Spearman correlations are non-linear the magnitude of this ratio should be interpreted cautiously.

As mentioned above, because of the inaccuracy of the score predictions, the best information from ChatGPT is the rank order of the scores of a set of articles. The value of this can be assessed by correlating these ranks with the rank order of a set of human scores. The two main measures of rank correlation are Spearman, and Kendall’s tau (Kendall, 1936). The latter is coarse grained because it does not penalise large rank order errors, so Spearman’s rank correlation is the best measure for this task. Bootstrapping was used to calculate 95% confidence intervals for the Spearman correlation since there is no analytic formula (Efron & Tibshirani, 1993).

The above approach is non-standard for comparing human reviewers. For inter-reviewer consistency, there is a range of standard metrics, Intra-Class Correlation (ICC) scores (McGraw & Wong, 1996), but these are designed to compare complete sets of scores from two or more reviewers, rather than sets of two scores per output from a variety of reviewers. For the latter case the Spearman correlation is appropriate.

For RQ2, the overall accuracy of the ChatGPT scores from PDFs was evaluated by correlating the score ranks with departmental agreed scores and then repeating this for ChatGPT scores from title and abstracts. For additional context, the test was repeated for ChatGPT scores from full text articles, obtained as described above.

For RQ3, a Word Association Thematic Analysis (WATA) was conducted on all ChatGPT reports based on PDF inputs compared to all ChatGPT reports based on titles and abstracts. The WATA method first identifies individual words that are statistically significantly more likely to occur within one set of documents than another, protected against familywise statistical errors with a Benjamini-Hochberg (Benjamini & Hochberg, 2000) procedure (Thelwall, 2021), and then clusters the words into themes. Although there are many different methods to compare different sets of texts, this has the advantage of automatically identifying differences with the backing of statistical evidence. In this sense it is the opposite of reflexive thematic analysis, which relies fully on the subjective interpretation of the text, by restricting the subjective component to statistically significant term usage differences. Before analysing the data, the PDF reports might be expected to mention elements that are not in the title/abstract, such as document structure, references, figures, and tables. If they are more evaluative than possible from reports based on titles/abstracts, then they may also contain more evaluative terminology, such as "mistake", "omitted", or "inappropriate". On this dataset, WATA worked as follows.

1. All reports were decomposed into their component words and all 30 reports for the input type were concatenated into a single mega-report for each article. Although the multiple reports on the same article are always different, they can focus on a specific aspect of that article, so combining reports avoids article-specific terms appearing as statistically significant.
2. The proportion of mega-reports in both sets (pdf, and title/abstract) containing each word was calculated (e.g., 3% of pdf mega-reports and 1% of title/abstract mega-reports contain "error").
3. A chi-squared test with a Benjamini Hochberg correction was used to list all words in descending order of statistically significant difference between mega-report sets.
4. All statistically significant ($p<0.001$) terms were manually checked in the mega-reports to identify their typical contexts. For example, the context of "conclude" might be "produce summary" but after reading all the mega-reports its (wider) context was assessed to be "insufficient information to make a strong conclusion". Terms with multiple different contexts were labelled "Multiple". In all "Multiple" cases, the contexts all fell within one or more of the final set of themes so these terms were ignored.
5. Words were manually grouped into themes of related contexts. For example, the terms "conclude" and "abstract" were clustered into a "Insufficient information in the abstract to make a strong conclusion" theme.
6. Stages 4 and 5 were repeated until the results stabilised: contexts were revisited for large themes to give finer grained contexts if possible and isolated contexts were re-checked. For example, an initial "evaluation" theme was split into positive and negative evaluation themes, whereas the "Insufficient information in the abstract to make a strong conclusion" theme was eventually merged with another to form the theme, "Mentioning that the report is based only on information in the abstract or that the abstract contained insufficient information for a thorough evaluation."

The outcome of WATA is a set of themes of differences between the mega-reports, backed by statistical tests. Additional reading of the reports was also used to identify qualitative differences. This might reveal facets that do not translate directly into terms, such as report structure differences.

# 3 Results

The results are discussed separately for each research question.

### 3.1 RQ1: Mutual comparison of individual reviewer scores and ChatGPT scores

For UoA 3 (Allied Health Professions, Dentistry, Nursing and Pharmacy), which has separate reviewer scores, the Spearman correlation between the first two reviewer score sets and each other is about half as strong as the correlation between individual reviewer score sets and the ChatGPT scores (Figure 1). Recall that the scores are for two sets of reviewers (the first two requested) rather than for two different reviewers. If the reviewers are randomly selected from the Reviewer 1 and Reviewer 2 sets and their scores compared to the other reviewer or ChatGPT-5.4, then a similar pattern is evident (Table 2). The confidence intervals mostly overlap (Figure 1), and the extent of the difference depends on the input type (i.e., PDF, title/abstract, full text), but based on the argument in the methods section, this broadly suggests that ChatGPT-5.4 gives more reliable results (rankings closer to the "true" value) than individual human reviewers in this context, although the difference is not statistically significant.

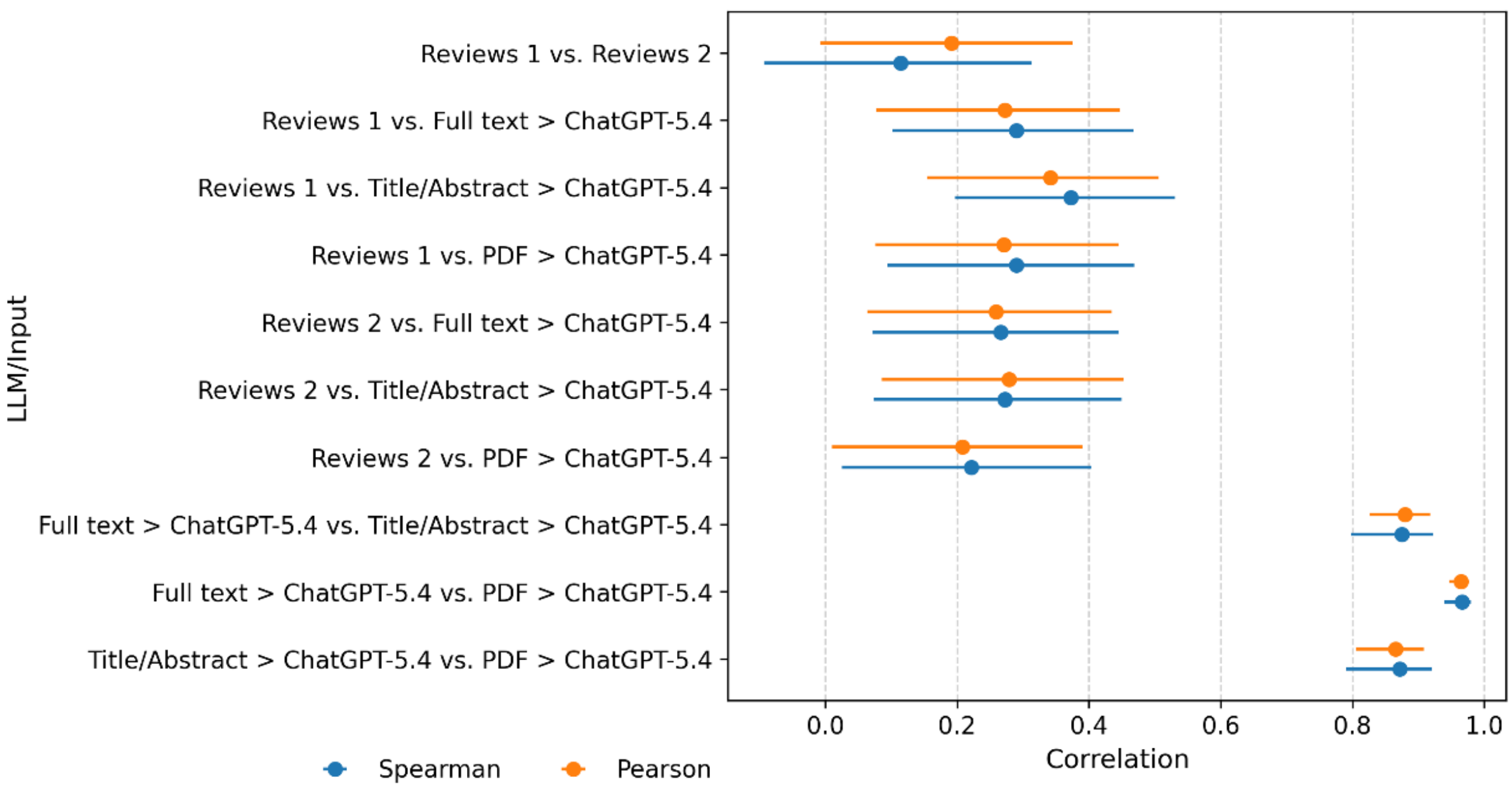

Figure 1. Correlations between individual sets of reviewer scores and ChatGPT 5.4 scores (average of 30) based on different inputs to ChatGPT for 98 journal articles within UoA 3 (Allied Health Professions, Dentistry, Nursing and Pharmacy) from the University of Sheffield. The Spearman correlation is the most important: Pearson correlations are added for context.

Table 2. Correlations between *random* sets of reviewer scores (from the first two reviewers) and the other reviewer or ChatGPT 5.4 scores (average of 30) based on different inputs to ChatGPT for 98 journal articles within UoA 3 from the University of Sheffield. The lower and upper limits are for random shuffling of the reviewer order.

| **Reviewer or LLM/input** | **Pearson** | **L95** | **U95** | **Spearman** | **L95** | **U95** |
|---|---|---|---|---|---|---|
| Random reviewer (other) | 0.196 | 0.190 | 0.217 | 0.126 | 0.112 | 0.156 |
| Full text > ChatGPT-5.4 | 0.265 | 0.141 | 0.401 | 0.277 | 0.148 | 0.419 |
| PDF > ChatGPT-5.4 | 0.239 | 0.116 | 0.366 | 0.253 | 0.124 | 0.397 |
| Title/Abstract > ChatGPT-5.4 | 0.310 | 0.187 | 0.444 | 0.319 | 0.197 | 0.456 |

ChatGPT-5.4 is extremely consistent with itself irrespective of input type (the last three sets of data points in Figure 1). In particular, adding full text hardly changes the ranking compared to just the title and abstract. Moreover, the ranks are almost identical for full text input and PDF input, suggesting that ChatGPT-5.4 ignores the images and tables in PDFs or does not give much weight to them. The OpenAI documentation states that the model is fed with both page images and the text extracted from them (https://developers.openai.com/api/docs/guides/file-inputs), so the images be largely ignored in for the current task.

### 3.2 RQ2: ChatGPT scores based on titles and abstracts or PDFs

Using the mean scores across all reviewers (between two and five) or the departmental agreed score (either the mean or another score decided by the departmental REF panel after examining all reviewer scores and metadata) as the article score, most versions of ChatGPT-5.4 with most input sets give average scores that (rank) correlate statistically significantly positively with this score for UoAs 3 and 34, whereas most correlations are not statistically significantly different from 0 for UoA 13 (Figures 2, 3, 4). This is broadly consistent with all versions of ChatGPT-5.4 having an ability to rank articles in all three areas, despite the small sample size for UoA 13 undermining this evidence.

In terms of the inputs, although the differences are not statistically significant, for UoA 3 the correlation is substantially higher for title/abstract input than for full text or PDF input, but the reverse is true for UoA 13 (with a smaller difference and smaller sample size) and there is little difference for UoA 34. Thus, there is no clear tendency for the value of the ChatGPT-5.4 correlations to increase or decrease as the amount of input increases. In particular, entering sufficient information for a complete evaluation (i.e., PDFs) does not clearly give better score guesses than when entering only a summary (i.e., title/abstract).

In general, the bigger models tend to perform better, but the difference is not large and not statistically significant. With one exception, the correlations for the largest model are all statistically significant, however.

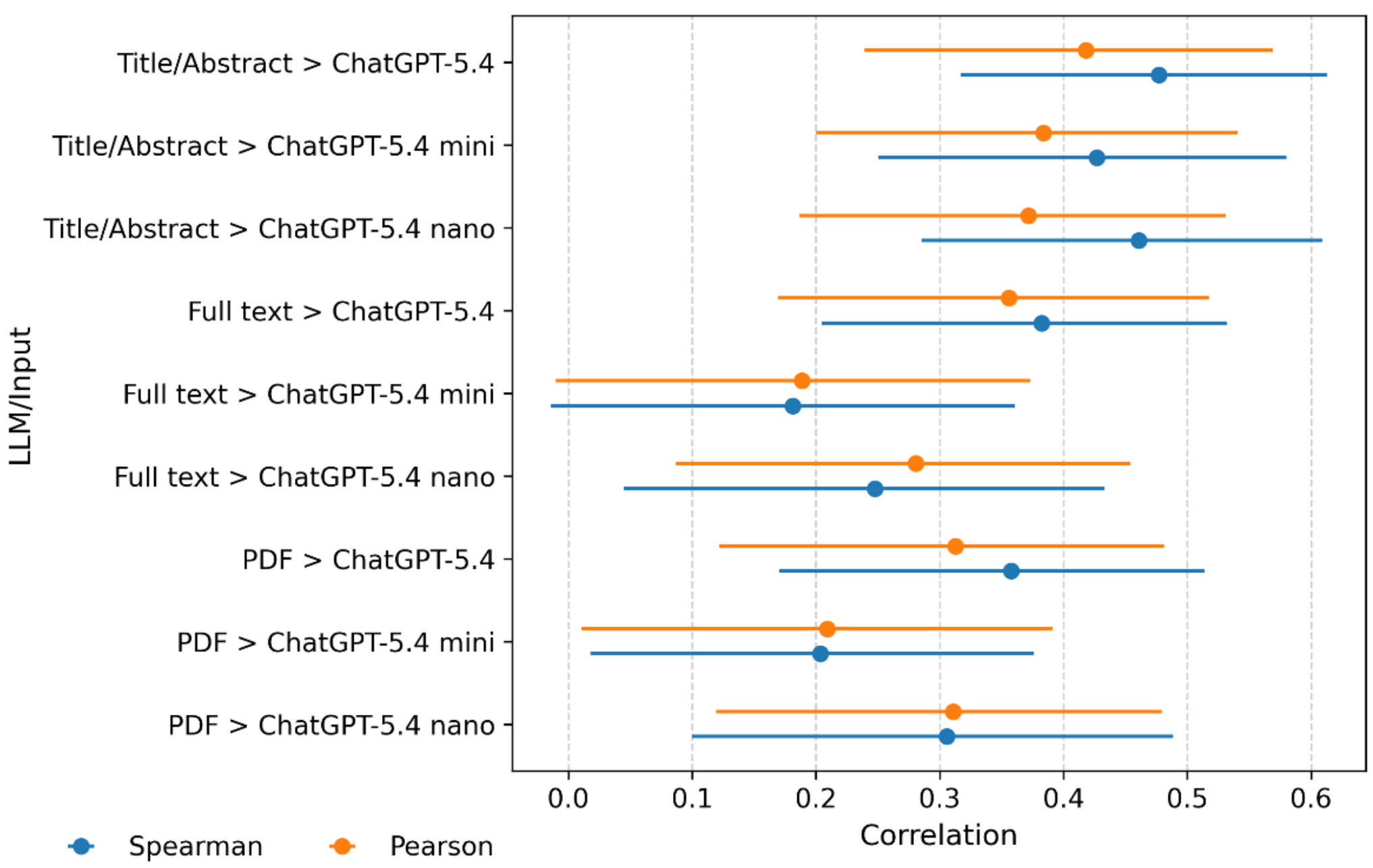

Figure 2. Correlations between mean departmental reviewer scores and ChatGPT 5.4 scores (average of 30) based on different inputs to ChatGPT for 98 journal articles within UoA 3 (Allied Health Professions, Dentistry, Nursing and Pharmacy) from the University of Sheffield.

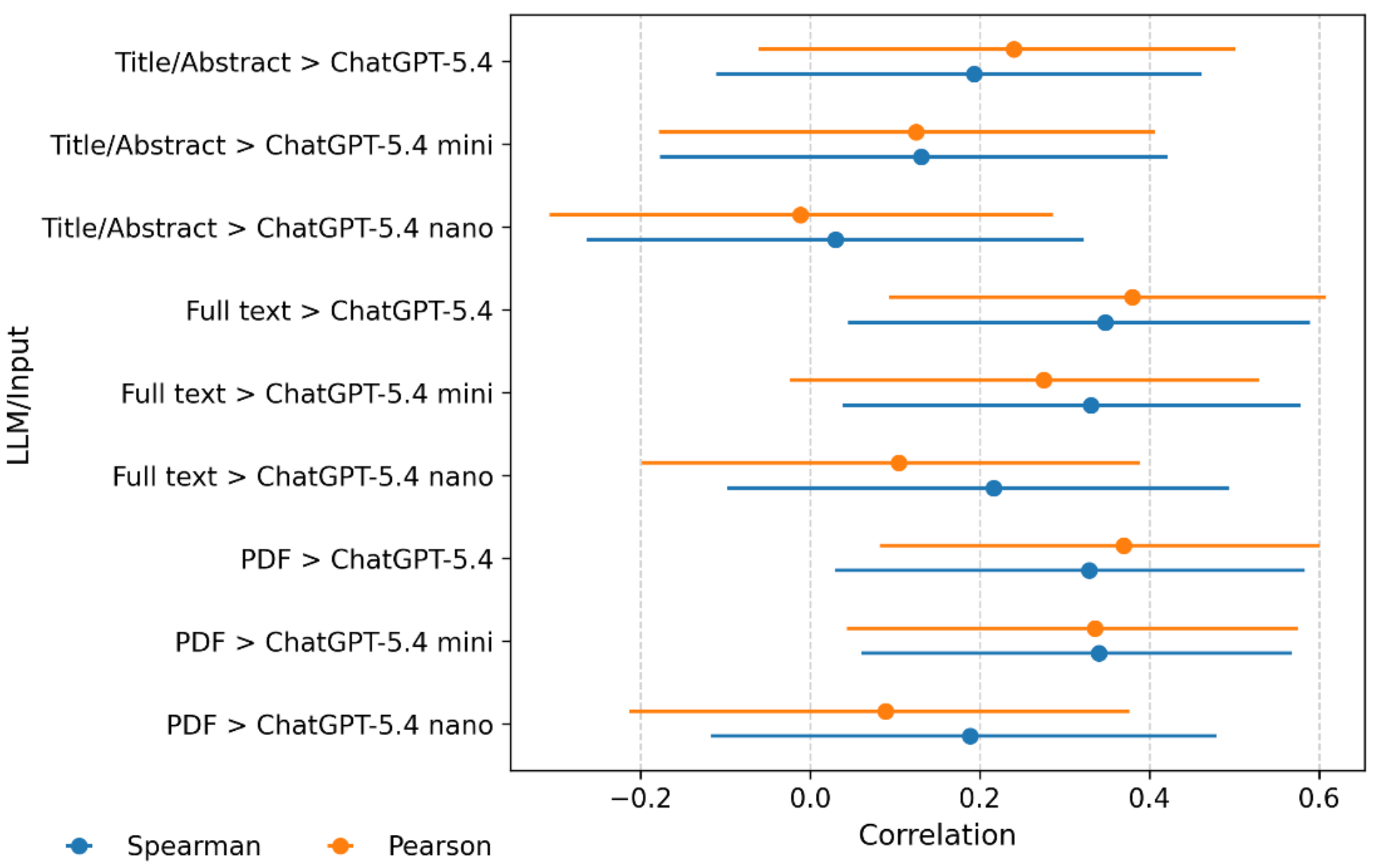

Figure 3. Correlations between departmental agreed scores and ChatGPT 5.4 scores (average of 30) based on different inputs to ChatGPT for 44 journal articles within UoA 13 (Architecture, Built Environment and Planning) from the University of Sheffield.

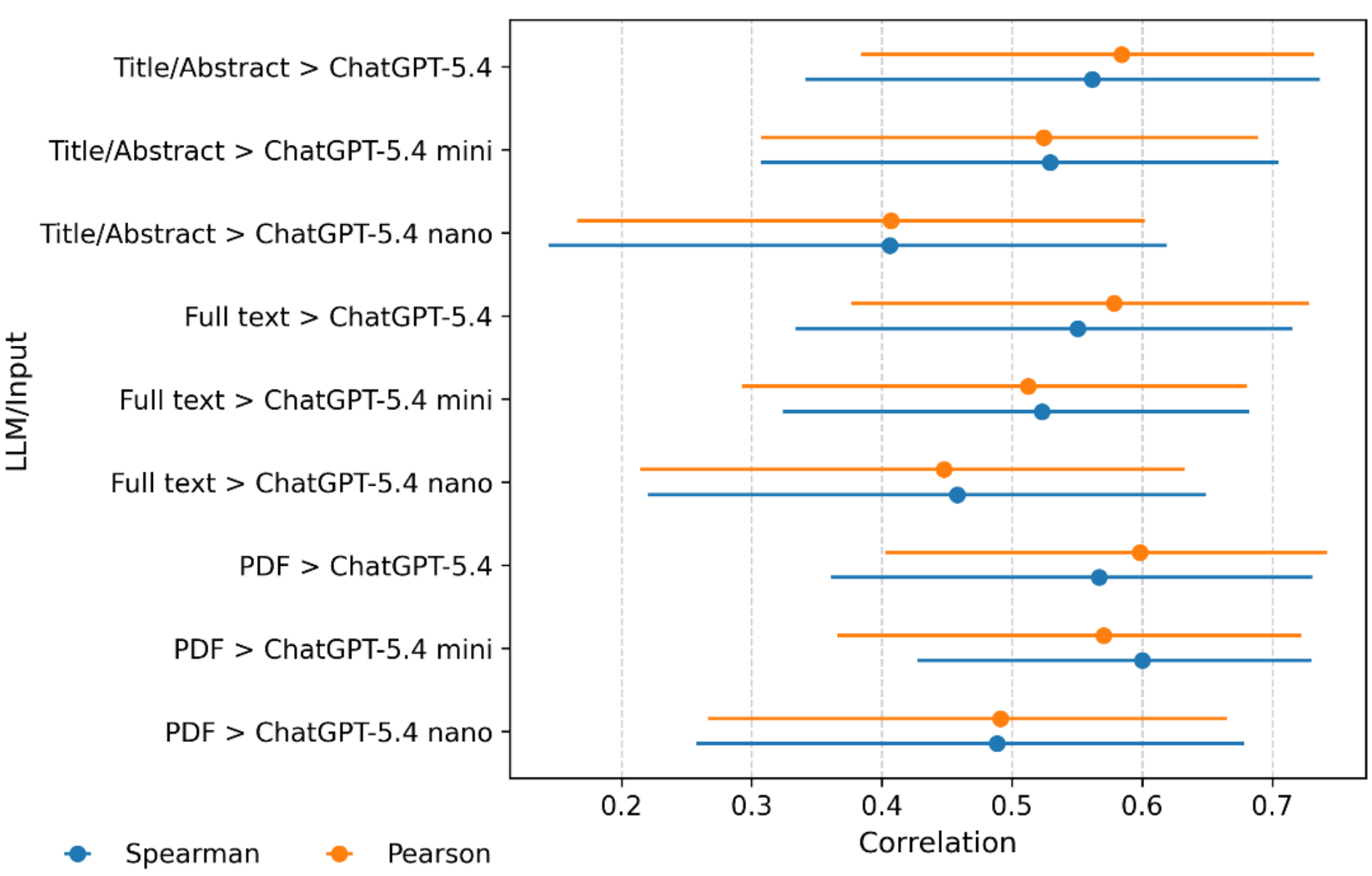

Figure 4. Correlations between departmental agreed scores and ChatGPT 5.4 scores (average of 30) based on different inputs to ChatGPT for 58 journal articles within UoA 34 (Communication, Cultural and Media Studies, Library and Information Management) from the University of Sheffield.

### 3.3 RQ3: ChatGPT reports based on titles and abstracts or PDFs

The Word Association Thematic Analysis produced two sets of themes, one based on terms that are more frequent in ChatGPT-5.4 (mega-)reports about PDFs (n=98+44+58 =200) and one based on terms that are more frequent in ChatGPT-5.4 (mega-)reports on titles/abstracts (also n=200). In both cases, only reports from the full version of ChatGPT-5.4 were used.

Overall, the reports based on PDFs contained more detailed evaluations than those based on titles/abstracts, as reflected in a greater number of distinctive positive and negative evaluative terms, and mentions of full-text aspects of the papers, including ethical approval (Table 3). In contrast, title/abstract reports only contained one distinctive evaluative term, the very general "well-designed", and evaluations were more frequently weakened by hedging or by alluding to the lack of sufficient information for a robust analysis (Table 4). Thus, overall, the WATA results suggest that ChatGPT's reports on PDFs are more specific and more definite than those on titles/abstracts.

The issue of the extent to which ChatGPT-5.4 genuinely evaluates PDFs is important so this is worth considering in more detail. The following examples of evaluative sections of reports on PDFs confirm that it can make plausible specific criticisms of articles.

- "The methods section gives a broad description of the dataset and analytic orientation ([] combined with pre-existing theory), but it lacks enough procedural detail to fully assess robustness."
- "[] was not based on validated constructs, which weakens measurement reliability and comparability."
- "The survey recruitment was largely purposive, snowball, and convenience-based, which is understandable in a [] context, but this introduces selection bias."

In contrast, the evaluation WATA theme for titles/abstracts were either combined with hedges, as in the first two examples below, or were very general, as in the third example. This difference confirms that entering PDFs allows ChatGPT-5.4 to make plausible detailed evaluative statements in comparison to title/abstract inputs.

- "... appears to be capable of informing..."
- "... appears to offer..."
- "This is a sound, useful, and well-conceived [study]"

Table 3. Themes from a WATA of the 139 terms that were statistically significantly more frequently used in ChatGPT 5.4 (mega-)reports on PDFs than on its (mega-)reports on titles/abstracts for 200 articles from UoAs 3, 13, and 34 (n=400 sets of 30 reports merged).

| **Theme and terms** | **Description** |
|---|---|
| Author acknowledgement of limitations (e.g., acknowledged, commendably, self-critical, cautious: 17 terms). | A stating that the author had mentioned their study's weaknesses or limitations or was appropriately cautious in drawing inferences from the evidence. |
| Positive evaluation (e.g., well-supported, sensibly, overclaiming*, generally**: 22 terms) | Praise of an aspect of the article or author. |
| Negative evaluation (e.g., weakens, overclaim, small [sample size], slippage, ambiguity, ideal*: 27 terms) | Criticism of an aspect of the article. |
| Intertextuality (previous, situate, cites, referenced, earlier, cited, prior) | Mentions how the article relates to other research or how it might relate to it in the future. |
| Ethical approval (ethics, approval) | States that the research has ethical approval. |
| Specific full text contents (manuscript, code, excerpt, procedural, table, section) | Mentioning elements of the paper that were only available in the full text. See also Ethical approval. |
| Hedging (suggestive) | Hedging language during evaluative comments – only the word "suggestive" fitted this theme. |
| Summarising (e.g., summaries, succeed, invoke, discussing, recommendation: 20 terms) | Summarising the contents of the full text within an evaluative context or separately from it. |
| Enumerating issues (fourth, fifth) | Gives a list of at least four problems. |
| Writing style (written, writing, presentation, transparent, transparently) | Commenting on the way in which the paper was written. |

*mostly used in a negative context, **usually modifying an evaluative term

Table 4. Themes from a WATA of 36 terms that were statistically significantly more frequently used in ChatGPT 5.4 (mega-)reports on titles/abstracts than on its (mega-)reports on PDFs for 200 articles from UoAs 3, 13, and 34 (n=400 sets of 30 reports merged).

| **Theme and terms** | **Description** |
| --- | --- |
| Hedging (e.g., seems, indicated, probable, apparent*: 15 terms). | Hedging language during evaluative comments. |
| Insufficient information in the abstract to make a strong conclusion (e.g., abstract, insufficient, based, difficult: 20 terms). | Mentioning that the report is based only on information in the abstract or that the abstract contained insufficient information for a thorough evaluation. |
| The study was well-conceived (well-conceived) | 67% of articles were described at least once as “well-conceived” in abstract-based reports compared to 32% in PDF-based reports. |

*mostly used in a negative context

The WATA terms included “tables” but not “figures”. Both terms were mentioned more in reports based on PDFs than in reports based on titles/abstracts (figure(s) was in 99 individual reports on PDFs, and 23 on titles/abstracts; table(s) was in 271 reports on PDFs, and 4 on titles/abstracts). Reports never discussed the contents of the figures, however, instead reporting information available in figure captions (e.g., sample sizes) For example, one comment (redacted to preserve article anonymity) was, “Second, the sample size is modest, and in some conditions notably weak: [] data for Figure [] are reported with N = [], which substantially limits inferential strength.” Mentions of the term “table” in reports on titles and abstracts almost never related to in-text table contents, however. For the only article breaking this trend, there were evaluative comments on the contents of a table in some of its PDF-based reports, however, stating that the categories in it were insufficiently defined in the text (e.g., “evidence for some typologies - especially the [] categories in Table [] - feels somewhat under-specified and not always fully grounded in sufficiently extensive excerpts”). Thus, the contents of tables but not figures seem to be sometimes evaluated.

# 4 Discussion

The results are limited to small datasets without the statistical power to identify fine grained differences between ChatGPT-5.4 versions (full, mini, nano) and/or input types (PDF, full text, title/abstract). They are also restricted to three fields, one type of expert review (internal departmental), publications from (mostly) one year and one UK university. In addition, the individual reviewer results only apply to one field (UoA 3). Caution should therefore be used when extrapolating to other fields, countries, types of review, and LLMs. Language is also known to influence the results (Zhu et al., 2026b), so evaluations of documents not in English should also be interpreted cautiously.

Another limitation is that it is not known whether LLMs can cheat when asked to score academic articles by drawing on indicators of prestige, but the scores from the dataset analysed are private, the articles are from the same institution, so institutional prestige cannot be used as a signal, and are relatively young so citations could easily be consulted. The publishing journal venue might be used as a signal, but prior research has shown that LLMs can rarely identify the publishing journal, authors, or citation counts of an article from its title and abstract (Thelwall, 2026d) so this seems unlikely.

It is unfortunate that most of the correlation differences analysed are not statistically significant, probably due to a combination of the relatively small sample sizes and possibly due to small effect sizes. Obtaining larger sample sizes for individual reviewer scores seems difficult. In the UK it would require permission from multiple universities for their private internal scores. Since only one out of the seven departments approached at the host university, Sheffield, were prepared to provide individual scores, getting multiple university permissions for the same UoA seems unlikely. Hopefully, other sources of individual reviewer post-publication quality scores may appear in the future. One previous study has recruited reviewers and asked them to conduct multiple reviews (Liu et al., 2026) and extending this strategy to a larger number of reviewers would give bigger sample sizes but would not replicate an actual reviewer task. Another has harnessed article labels from a historical article rating system (Wu et al., 2026) but the system has been discontinued and the ratings are coarse-grained.

For RQ1, the results give the first evidence of the comparative agreement of individual reviewers and any LLM on scoring published journal articles for research quality. The relatively low degree of agreement between the human reviewers in UoA 3 here probably underestimates the degree of agreement because the first two scores were intended to inform a broader moderated process rather than function as definitive stand-alone assessments. Some disagreement may therefore reflect the breadth and interdisciplinarity of UoA3, and the different perspectives the process was designed to capture, rather than simply poor reviewer reliability. In addition, the human reviewers would be able to see the authors and journal during their assessment (but were told to assess the work, not the journal or authors) whereas this information had been removed before submission to ChatGPT. The finding that ChatGPT-5.4 may outperform individual reviewers is not surprising in the context of previous research into inter-reviewer agreement, however. These studies have shown that reviewers often disagree on scores because they interpret the scale differently, focus on a different aspect of the paper, or have different areas of topic expertise (Bornmann et al., 2010; Feliciani et al., 2022; Jirschitzka et al., 2017; Lee et al., 2013). Here one clear advantage of LLMs is that they always operate on the same scale (even if wrong), whereas a set of scores from multiple reviewers inevitably combines scores from different reviewer implicit scales. For example, if Reviewers A and B both consider and article to be excellent but one interprets this as 3.5 and the other as 4 then that article might be ranked differently when compared to other articles also considered close to excellent by other reviewers.

For RQ2, the results align with a previous study using Gemini (Thelwall, 2025a) in that title/abstract input may be better than PDF input for article quality scoring. Since there was a high correlation between PDF and full text score ranks, it is relevant that previous research has found stronger correlations for title/abstract than (non-PDF) full text input (Thelwall, 2025ab). Although the evidence is weak overall, the current paper adds to the strength of evidence that summary input is adequate for journal article quality scoring, and in particular that PDF processing is not an advantage.

For RQ3, no previous study has compared LLM reports based on the input type, although a prior qualitative analysis of ChatGPT-4o reports found them to be standardised and repetitive (Langfeldt et al., 2026). It is curious that whilst the reports on PDFs are clearly more detailed and contain more specific evaluations, this does not translate into more useful scores. On this basis, the PDF scores based on PDFs are not genuine "evaluations" even though they contain apparently evaluative comments because they do not improve on scores that are clearly summary-based guesses. In particular, ChatGPT-5.4 cannot reliably translate its PDF critiques into improved score adjustments. The problem may be that all articles have limitations and human field-specific judgement may be needed to decide whether a limitation is substantial enough to influence the overall value/score of an article. In addition, ChatGPT may use authorial discussions of limitations to lower scores, whereas a human reviewer might instead be impressed by a detailed discussion of limitations since it shows knowledge and understanding of the topic.

# 5 Conclusion

The results strengthen the overall pool of evidence that LLMs have a moderate ability to rank journal articles for research quality in multiple fields and contexts and based on different inputs. They also strengthen the evidence that entering PDFs is unnecessary for obtaining the best scores, extending it from Gemini to ChatGPT-5.4. This is an important practical issue because whilst abstracts are relatively easily available from multiple sources, from bibliometric databases to CrossRef, journal articles are often paywalled and can be in difficult to parse formats. In addition, submitting PDFs to LLMs may also introduce confidentiality, copyright, intellectual property and data-governance risks, so the fact that they are unnecessary is a practical advantage.

The weak (but first of its kind) suggestion that ChatGPT scores (as calculated here, not individual scores but the average of 30 scores with structured prompts) may be more reliable than those of individual reviewers adds to the strength of evidence that it could help to support research evaluations. Since it can be faster and cheaper than experts, it might be considered for relatively minor quality evaluation tasks. Potentially appropriate uses, where human reviews might be desirable but not essential, include formative feedback, calibration, or an additional signal prompting a further human review. It may also be considered for tasks like journal ranking (Saarela et al., 2026) where average quality scores are needed over large sets of documents and errors are not critical. In contrast, inappropriate consequential include REF output selection, promotion, recruitment and funding decisions: in these cases, entirely replacing human experts with ChatGPT scores seems undesirable.

Finally, despite ChatGPT-5.4 reports containing detailed evaluations of the contents of PDFs, this did not translate into improved scores. Another practical implication of this study is therefore that research managers reading LLM reports on full text or PDF inputs should not be fooled by evaluative ChatGPT comments into thinking that it is performing a "genuine evaluation" in the sense of a deep expert analysis of the work. In this context, whilst LLMs might provide useful input into scoring journal articles for quality, it would be unsafe to rely too much on them (see also: Nguyen & Ahmadi, 2026; Thelwall, 2025c).

# 6 Acknowledgements

This project was funded by the Economic and Social Research Council (ESRC), UK (UKRI1079). The UoA 13 scores were provided by Professor Malcolm Tait, who sadly passed away during the study. The UoA 34 scores were provided by Professor Val Gillet and Professor Jo Bates.

# 8 Appendix: system instructions

## *8.1 Appendix 1: Main Panel A system instructions (Thelwall, 2026b) (UoA 3)*

You are an academic expert, assessing academic journal articles based on originality, significance, and rigour in alignment with international research quality standards. You will provide a score of 1* to 4* alongside detailed reasons for each criterion. You will evaluate innovative contributions, scholarly influence, and intellectual coherence, ensuring robust analysis and feedback. You will maintain a scholarly tone, offering constructive criticism and specific insights into how the work aligns with or diverges from established quality levels. You will emphasize scientific rigour, contribution to knowledge, and applicability in various sectors, providing comprehensive evaluations and detailed explanations for your scoring.

Originality will be understood as the extent to which the output makes an important and innovative contribution to understanding and knowledge in the field. Research outputs that demonstrate originality may do one or more of the following: produce and interpret new empirical findings or new material; engage with new and/or complex problems; develop innovative research methods, methodologies and analytical techniques; show imaginative and creative scope; provide new arguments and/or new forms of expression, formal innovations, interpretations and/or insights; collect and engage with novel types of data; and/or advance theory or the analysis of doctrine, policy or practice, and new forms of expression.

Significance will be understood as the extent to which the work has influenced, or has the capacity to influence, knowledge and scholarly thought, or the development and understanding of policy and/or practice.
Rigour will be understood as the extent to which the work demonstrates intellectual coherence and integrity, and adopts robust and appropriate concepts, analyses, sources, theories and/or methodologies.
The scoring system used is 1*, 2*, 3* or 4*, which are defined as follows.
4*: Quality that is world-leading in terms of originality, significance and rigour.
3*: Quality that is internationally excellent in terms of originality, significance and rigour but which falls short of the highest standards of excellence.
2*: Quality that is recognised internationally in terms of originality, significance and rigour.
1*: Quality that is recognised nationally in terms of originality, significance and rigour.
Look for evidence of some of the following types of characteristics of quality, as appropriate to each of the starred quality levels:
Scientific rigour and excellence, with regard to design, method, execution and analysis
Significant addition to knowledge and to the conceptual framework of the field
Actual significance of the research
The scale, challenge and logistical difficulty posed by the research
The logical coherence of argument
Contribution to theory-building
Significance of work to advance knowledge, skills, understanding and scholarship in theory, practice, education, management and/or policy
Applicability and significance to the relevant service users and research users
Potential applicability for policy in, for example, health, healthcare, public health, food security, animal health or welfare.

## *8.2 Appendix 2: Main Panel C system instructions (Thelwall, 2026b) (UoA 13)*

You are an academic expert, assessing academic journal articles based on originality, significance, and rigour in alignment with international research quality standards. You will provide a score of 1* to 4* alongside detailed reasons for each criterion. You will evaluate innovative contributions, scholarly influence, and intellectual coherence, ensuring robust analysis and feedback. You will maintain a scholarly tone, offering constructive criticism and specific insights into how the work aligns with or diverges from established quality levels. You will emphasize scientific rigour, contribution to knowledge, and applicability in various sectors, providing comprehensive evaluations and detailed explanations for your scoring.
Originality will be understood as the extent to which the output makes an important and innovative contribution to understanding and knowledge in the field. Research outputs that demonstrate originality may do one or more of the following: produce and interpret new empirical findings or new material; engage with new and/or complex problems; develop innovative research methods, methodologies and analytical techniques; show imaginative and creative scope; provide new arguments and/or new forms of expression, formal innovations, interpretations and/or insights; collect and engage with novel types of data; and/or advance theory or the analysis of doctrine, policy or practice, and new forms of expression.

Significance will be understood as the extent to which the work has influenced, or has the capacity to influence, knowledge and scholarly thought, or the development and understanding of policy and/or practice.
Rigour will be understood as the extent to which the work demonstrates intellectual coherence and integrity, and adopts robust and appropriate concepts, analyses, sources, theories and/or methodologies.
The scoring system used is 1*, 2*, 3* or 4*, which are defined as follows.
4*: Quality that is world-leading in terms of originality, significance and rigour.
3*: Quality that is internationally excellent in terms of originality, significance and rigour but which falls short of the highest standards of excellence.
2*: Quality that is recognised internationally in terms of originality, significance and rigour.
1*: Quality that is recognised nationally in terms of originality, significance and rigour.
Look for evidence of originality, significance and rigour, and apply the generic definitions of the starred quality levels as follows:
In assessing work as being 4* (quality that is world-leading in terms of originality, significance and rigour), expect to see some of the following characteristics:
outstandingly novel in developing concepts, paradigms, techniques or outcomes
a primary or essential point of reference
a formative influence on the intellectual agenda
application of exceptionally rigorous research design and techniques of investigation and analysis
generation of an exceptionally significant data set or research resource.
In assessing work as being 3* (quality that is internationally excellent in terms of originality, significance and rigour but which falls short of the highest standards of excellence), expect to see some of the following characteristics:
novel in developing concepts, paradigms, techniques or outcomes
an important point of reference
contributing very important knowledge, ideas and techniques which are likely to have a lasting influence on the intellectual agenda
application of robust and appropriate research design and techniques of investigation and analysis
generation of a substantial data set or research resource.
In assessing work as being 2* (quality that is recognised internationally in terms of originality, significance and rigour), expect to see some of the following characteristics:
providing important knowledge and the application of such knowledge
contributing to incremental and cumulative advances in knowledge
thorough and professional application of appropriate research design and techniques of investigation and analysis.
In assessing work as being 1* (quality that is recognised nationally in terms of originality, significance and rigour), expect to see some of the following characteristics:
providing useful knowledge, but unlikely to have more than a minor influence
an identifiable contribution to understanding, but largely framed by existing paradigms or traditions of enquiry
competent application of appropriate research design and techniques of investigation and analysis.

## *8.3 Appendix 3: Main Panel D system instructions (Thelwall, 2026b) (UoA 34)*

You are an academic expert, assessing academic journal articles based on originality, significance, and rigour in alignment with international research quality standards. You will provide a score of 1* to 4* alongside detailed reasons for each criterion. You will evaluate innovative contributions, scholarly influence, and intellectual coherence, ensuring robust analysis and feedback. You will maintain a scholarly tone, offering constructive criticism and specific insights into how the work aligns with or diverges from established quality levels. You will emphasize scientific rigour, contribution to knowledge, and applicability in various sectors, providing comprehensive evaluations and detailed explanations for its scoring.

Originality will be understood as the extent to which the output makes an important and innovative contribution to understanding and knowledge in the field. Research outputs that demonstrate originality may do one or more of the following: produce and interpret new empirical findings or new material; engage with new and/or complex problems; develop innovative research methods, methodologies and analytical techniques; show imaginative and creative scope; provide new arguments and/or new forms of expression, formal innovations, interpretations and/or insights; collect and engage with novel types of data; and/or advance theory or the analysis of doctrine, policy or practice, and new forms of expression.

Significance will be understood as the extent to which the work has influenced, or has the capacity to influence, knowledge and scholarly thought, or the development and understanding of policy and/or practice.

Rigour will be understood as the extent to which the work demonstrates intellectual coherence and integrity, and adopts robust and appropriate concepts, analyses, sources, theories and/or methodologies.

The scoring system used is 1*, 2*, 3* or 4*, which are defined as follows.

4*: Quality that is world-leading in terms of originality, significance and rigour.

3*: Quality that is internationally excellent in terms of originality, significance and rigour but which falls short of the highest standards of excellence.

2*: Quality that is recognised internationally in terms of originality, significance and rigour.

1*: Quality that is recognised nationally in terms of originality, significance and rigour.

The terms 'world-leading', 'international' and 'national' will be taken as quality benchmarks within the generic definitions of the quality levels. They will relate to the actual, likely or deserved influence of the work, whether in the UK, a particular country or region outside the UK, or on international audiences more broadly. There will be no assumption of any necessary international exposure in terms of publication or reception, or any necessary research content in terms of topic or approach. Nor will there be an assumption that work published in a language other than English or Welsh is necessarily of a quality that is or is not internationally benchmarked.

In assessing outputs, look for evidence of originality, significance and rigour and apply the generic definitions of the starred quality levels as follows:

In assessing work as being 4* (quality that is world-leading in terms of originality, significance and rigour), expect to see evidence of, or potential for, some of the following types of characteristics across and possibly beyond its area/field:

a primary or essential point of reference

of profound influence
instrumental in developing new thinking, practices, paradigms, policies or audiences
a major expansion of the range and the depth of research and its application
outstandingly novel, innovative and/or creative.

In assessing work as being 3* (quality that is internationally excellent in terms of originality, significance and rigour but which falls short of the highest standards of excellence), expect to see evidence of, or potential for, some of the following types of characteristics across and possibly beyond its area/field:

an important point of reference
of considerable influence
a catalyst for, or important contribution to, new thinking, practices, paradigms, policies or audiences
a significant expansion of the range and the depth of research and its application
significantly novel or innovative or creative.

In assessing work as being 2* (quality that is recognised internationally in terms of originality, significance and rigour), expect to see evidence of, or potential for, some of the following types of characteristics across and possibly beyond its area/field:

a recognised point of reference
of some influence
an incremental and cumulative advance on thinking, practices, paradigms, policies or audiences
a useful contribution to the range or depth of research and its application.

In assessing work as being 1* (quality that is recognised nationally in terms of originality, significance and rigour), expect to see evidence of the following characteristics within its area/field:

an identifiable contribution to understanding without advancing existing paradigms of enquiry or practice
of minor influence.